\documentclass[prl,twocolumn,showpacs,aps,10pt,nofootinbib]{revtex4-1}
\usepackage{graphicx}
\usepackage{hyperref}
\usepackage{amsmath}
\usepackage{amsthm}
\usepackage{amssymb}
\usepackage{dsfont}
\usepackage{bbold}
\usepackage{comment}
\usepackage{color}
\newtheorem{theorem}{Theorem}

\newcommand{\proofend}{\hfill\fbox\\\medskip }

\newcommand{\bit}{\begin{itemize}}
\newcommand{\eit}{\end{itemize}\par\noindent}
\newcommand{\ben}{\begin{enumerate}}
\newcommand{\een}{\end{enumerate}\par\noindent}
\newcommand{\beq}{\begin{equation}}
\newcommand{\eeq}{\end{equation}\par\noindent}
\newcommand{\beqa}{\begin{eqnarray*}}
\newcommand{\eeqa}{\end{eqnarray*}\par\noindent}
\newcommand{\beqn}{\begin{eqnarray}}
\newcommand{\eeqn}{\end{eqnarray}\par\noindent}

\begin{document}

\title{A note on the joint measurability of POVMs and its implications for contextuality}
\author{Ravi Kunjwal}
\email{rkunj@imsc.res.in}
\affiliation{Optics \& Quantum Information Group, The Institute of Mathematical Sciences, C.I.T Campus, Tharamani, Chennai 600 113, India}

\date{\today}                                           

\begin{abstract}
The purpose of this note is to clarify the logical relationship between joint measurability and contextuality for quantum observables in view of recent developments \cite{LSW,KG,KHF,FyuOh}.
\end{abstract}

\pacs{03.65.Ta, 03.65.Ud}

\maketitle

\section*{Introduction}

In a recent work \cite{KG}, a new proof of contextuality---in the generalized sense of Spekkens \cite{Spe05, LSW}---was provided using positive operator-valued measures (POVMs) and the 
connection between joint measurability of POVMs and contextuality was explicated. It was later shown in \cite{KHF} that any joint measurability structure can be realized in quantum theory,
leaving open the question of whether contextuality can always be demonstrated in these joint measurability structures. Subsequent to these two developments, in Ref. \cite{FyuOh} a peculiar 
feature of POVMs with respect to joint measurability was pointed out: that there exist three measurements which are pairwise jointly measurable and triplewise jointly measurable but for 
which there exist pairwise joint measurements which do not admit a triplewise joint measurement. In this note, I will briefly put these results in context and point out the logical relationship 
between joint measurability and the possibility of contextuality. Also, throughout this note, `sharp measurement' will be synonymous with projection-valued measures (PVMs) and `unsharp measurement'
will be synonymous with POVMs that are not PVMs.
\section*{Uniqueness of joint measurement for Projection-Valued Measures}
Since the peculiarity of positive-operator valued measures (POVMs) in cases of interest here arises from the nonuniqueness of joint measurements, I will first prove the uniqueness of joint measurements for projection-valued measures (PVMs).
This will help clarify how the distinction between sharp and unsharp measurements comes to play a role in Specker's scenario \cite{KG}. 

Consider a nonempty set $\Omega_i$ and a $\sigma$-algebra $\mathcal{F}_i$ of subsets of $\Omega_i$, for $i\in\{1,\dots,N\}$. The POVM $M_i$ is defined as the map $M_i: \mathcal{F}_i\rightarrow \mathcal{B}_+(\mathcal{H})$,
where $\sum_{X_i\in\mathcal{F}_i}M_i(X_i)=I$ and $\mathcal{B}_+(\mathcal{H})$ denotes the set of positive semidefinite operators on a Hilbert space $\mathcal{H}$. $I$ is the identity operator
on $\mathcal{H}$. Therefore: $M_i\equiv\{M_i(X_i)|X_i\in\mathcal{F}_i\}$, where $X_i$ labels the elements of POVM $M_i$. $M_i$ becomes a projection-valued measure (PVM) under the 
additional constraint $M_i(X_i)^2=M_i(X_i)$ for all $X_i\in \mathcal{F}_i$.

\begin{theorem}\label{uniqueness}
Given a set of POVMs, $\{M_1,\dots,M_N\}$, all of which except at most one---say $M_N$---are PVMs, so that for $i\in\{1,\dots,N-1\}$
$$M_i\equiv\{M_i(X_i)|X_i\in\mathcal{F}_i, M_i(X_i)^2=M_i(X_i)\}$$ and $$M_N\equiv\{M_N(X_N)|X_N\in\mathcal{F}_N\},$$
the set of POVMs, $\{M_1,\dots,M_N\}$, is jointly measurable if and only if they commute pairwise, i.e.,
$$M_j(X_j)M_k(X_k)=M_k(X_k)M_j(X_j),$$
for all $j,k\in\{1,\dots,N\}$ and $X_j\in\mathcal{F}_j, X_k\in\mathcal{F}_k$. In this case, there exists a unique joint POVM $M$, 
defined as a map $$M:\mathcal{F}_1\times\mathcal{F}_2\times\dots\times\mathcal{F}_N \rightarrow \mathcal{B}_+(\mathcal{H}),$$ such that
$$M(X_1\times\dots\times X_N)=M_1(X_1)M_2(X_2)\dots M_N(X_N),$$
for all $X_1\times\dots\times X_N \in\mathcal{F}_1\times\dots\times \mathcal{F}_N.$
\end{theorem}

\emph{Proof.}---This proof is adapted from, and is a generalization of, the proof of Proposition 8 in the Appendix of Ref. \cite{heinosaari}. 

The first part of the proof is for the implication: joint measurability $\Rightarrow$ pairwise commutativity---A joint POVM for $\{M_1,\dots,M_N\}$ is defined as a map
$M:\mathcal{F}_1\times\mathcal{F}_2\times\dots\times\mathcal{F}_N \rightarrow \mathcal{B}_+(\mathcal{H})$, such that
\begin{equation}
 M_i(X_i)=\sum_{\{X_j\in\mathcal{F}_j|j\neq i\}}M(X_1\times\dots\times X_N)
\end{equation}
for all $X_i\in\mathcal{F}_i$, $i\in\{1\dots N\}$. Also, $M(X_1\times\dots\times X_N)\leq M_1(X_1)$, so the range of $M(X_1\times\dots\times X_N)$ is contained in the range of 
$M_1(X_1)$, and therefore:
\begin{equation}
M_1(X_1)M(X_1\times\dots\times X_N)=M(X_1\times\dots\times X_N).
\end{equation}
Using this relation for the complement $\Omega_1\backslash X_1 \in \mathcal{F}_1$:
\begin{eqnarray}
&&M_1(X_1)M(\Omega_1\backslash X_1\times\dots\times X_N)\nonumber\\
&&=(I-M_1(\Omega_1\backslash X_1))M(\Omega_1\backslash X_1\times\dots\times X_N)\nonumber\\
&&=0.
\end{eqnarray}
Taking the adjoints, it follows that
\begin{equation}
M(X_1\times\dots\times X_N)M_1(X_1)=M(X_1\times\dots\times X_N),
\end{equation}
and
\begin{equation}
M(\Omega_1\backslash X_1\times\dots\times X_N)M_1(X_1)=0.
\end{equation}
Denoting 
$$M^{(i)}(X_{i+1}\times\dots\times X_N)\equiv\sum_{\{X_j\in\mathcal{F}_j|j\leq i\}}M(X_1\times\dots\times X_N),$$
this implies:
\begin{eqnarray}
&&M_1(X_1)M^{(1)}(X_2\times\dots\times X_N)\nonumber\\
&=&M_1(X_1)M(X_1\times\dots\times X_N)\nonumber\\&&+M_1(X_1)M(\Omega_1\backslash X_1\times\dots\times X_N)\nonumber\\
&=&M_1(X_1)M(X_1\times\dots\times X_N)\nonumber\\
&=&M(X_1\times\dots\times X_N).
\end{eqnarray}
Taking the adjoint,
\begin{equation}
M^{(1)}(X_2\times\dots\times X_N)M_1(X_1)=M(X_1\times\dots\times X_N).
\end{equation}
Therefore:
\begin{eqnarray}
&&M_1(X_1)M^{(1)}(X_2\times\dots\times X_N)\nonumber\\
&=&M^{(1)}(X_2\times\dots\times X_N)M_1(X_1)\nonumber\\
&=&M(X_1\times\dots\times X_N).
\end{eqnarray}
Noting that $M^{(i-1)}(X_i\times\dots\times X_N)\leq M_i(X_i)$, one can repeat the above procedure for $M_i$, $i\in\{2,\dots,N-1\},$ to obtain:
\begin{eqnarray}
&&M^{(i-1)}(X_i\times\dots\times X_N)\nonumber\\
&=&M_i(X_i)M^{(i)}(X_{i+1}\times\dots\times X_N)\nonumber\\
&=&M^{(i)}(X_{i+1}\times\dots\times X_N)M_i(X_i).
\end{eqnarray}
Doing this recursively until $i=N-1$ and noting that $M^{(N-1)}(X_N)=M_N(X_N)$, it follows:
\begin{eqnarray}
&&M(X_1\times\dots\times X_N)\nonumber\\
&=&M_1(X_1)M^{(1)}(X_2\times\dots\times X_N)\nonumber\\
&=&M^{(1)}(X_2\times\dots\times X_N)M_1(X_1)\nonumber\\
&&\vdots\nonumber\\
&=&M_1(X_1)M_2(X_2)\dots M_{N-1}(X_{N-1})M_N(X_N)\nonumber\\
&=&M_N(X_N)M_{N-1}(X_{N-1})\dots M_2(X_2)M_1(X_1).\nonumber\\
\end{eqnarray}
For the last equality to hold, the POVM elements must commute pairwise, so that
\begin{equation}
M(X_1\times\dots\times X_N)=\prod_{i=1}^N M_i(X_i).
\end{equation}
This concludes the proof of the implication, joint measurability $\Rightarrow$ pairwise commutativity. The converse is easy to see since the joint POVM defined by taking the 
product of commuting POVM elements, $$\{M(X_1\times\dots\times X_N)=\prod_{i=1}^N M_i(X_i)|X_i\in\mathcal{F}_i\},$$ is indeed a valid POVM which coarse-grains to the given POVMs,
$\{M_1,\dots,M_N\}$.
\proofend

Indeed, pairwise commutativity $\Rightarrow$ joint measurability for any arbitrary set of POVMs, $\{M_1,\dots,M_N\}$, and it is only when all but (at most) one of these POVMs are PVMs that 
the converse---and the uniqueness of the joint POVM---holds.

\section*{Specker's scenario}
Specker's scenario requires a set of three POVMs, $\{M_1,M_2,M_3\}$, that are pairwise jointly measurable, i.e., $\exists$ POVMs $M_{12}$, $M_{23}$, and $M_{31}$ which measure the respective pairs jointly.
An immediate consequence of the requirement of pairwise joint measurability of $\{M_1,M_2,M_3\}$ is that in quantum theory these three measurements cannot be realized as projective measurements 
(PVMs) and still be expected to show any contextuality. This is because for projective measurements or projection-valued measures (PVMs), a set of three measurements that are pairwise jointly measurable---and therefore admit \emph{unique} pairwise joint measurements---are 
also triplewise jointly measurable in the sense that there exists a \emph{unique} triplewise joint measurement which coarse-grains to each pairwise implementation of the three measurements and therefore also to the single measurements.

From Theorem \ref{uniqueness}, it follows that if $M_i$, $i\in\{1,2,3\}$, are PVMs then they admit unique pairwise and triplewise joint PVMs: 
\begin{eqnarray}
M_{ij}(X_i\times X_j)&=&M_i(X_i)M_j(X_j),\\
M(X_1\times X_2\times X_3)&=&M_1(X_1)M_2(X_2)M_3(X_3),
\end{eqnarray}

corresponding to the maps $M_{ij}:\mathcal{F}_i\times\mathcal{F}_j\rightarrow \mathcal{B}_+(\mathcal{H})$ and $M:\mathcal{F}_1\times\mathcal{F}_2\times\mathcal{F}_3\rightarrow \mathcal{B}_+(\mathcal{H})$,
respectively. Intuitively, this is  easy to see since joint measurability is equivalent to pairwise commutativity for a set of projective measurements and the joint measurement for each pair is unique \cite{heinosaari}.
The existence of a unique joint measurement implies that there exists a joint probability distribution realizable via this joint measurement, thus explaining the pairwise statistics of the triple of measurements noncontextually in the traditional
Kochen-Specker sense.\footnote{KS-noncontextuality just means that there exists a joint probability distribution over the three measurement outcomes which marginalizes to the pairwise measurement statistics.
Violation of a KS inequality---obtained under the assumption that a global joint distribution exists---rules out KS-noncontextuality.}

Clearly, then, the three measurements $\{M_1, M_2, M_3\}$ must necessarily be unsharp for Specker's scenario to exhibit KS-contextuality. The uniqueness of joint measurements 
(pairwise or triplewise) need not hold in this case. I will refer to pairwise joint measurements as ``2-joints'' and triplewise joint measurements as ``3-joints''. Also,
I will use the phrases `joint measurability' and `compatibility' interchangeably since they will refer to the same notion. Consider 
the four propositions regarding the three measurements:

\begin{itemize}
 \item $\exists$ 2-joint: $\{M_1,M_2,M_3\}$ admit 2-joints,
 \item $\nexists$ 2-joint: $\{M_1,M_2,M_3\}$ do not admit 2-joints,
 \item $\exists$ 3-joint: $\{M_1,M_2,M_3\}$ admit a 3-joint,
 \item $\nexists$ 3-joint: $\{M_1,M_2,M_3\}$ do not admit a 3-joint,
\end{itemize}

The possible pairwise-triplewise propositions for the three measurements are: 
\begin{itemize}
 \item $(\exists \text{ 2-joint}, \exists \text{ 3-joint})$,
 \item $(\exists \text{ 2-joint}, \nexists \text{ 3-joint})$,
 \item $(\nexists \text{ 2-joint}, \nexists \text{ 3-joint})$.
\end{itemize}

Note that the proposition $(\nexists \text{ 2-joint}, \exists \text{ 3-joint})$ is trivially excluded because triplewise compatibility implies pairwise compatibility. Of the three remaining 
propositions, the ones of interest for contextuality are $(\exists \text{ 2-joint}, \exists \text{ 3-joint})$ and $(\exists \text{ 2-joint}, \nexists \text{ 3-joint})$,
since the remaining one is simply about observables that do not admit any joint measurement at all and hence no nontrivial measurement contexts exist for this proposition.\footnote{
It is worth noting that, if $\{M_1,M_2,M_3\}$ were PVMs, then there are only two possibilities: $(\exists \text{ 2-joint}, \exists \text{ 3-joint})$ and  $(\nexists \text{ 2-joint}, \nexists \text{ 3-joint})$,
since for three PVMs, $\exists \text{ 2-joint} \Leftrightarrow \exists \text{ 3-joint}$, because pairwise commutativity is equivalent to joint measurability and 
the joint measurements are unique on account of Theorem \ref{uniqueness}. This is why KS-contextuality is impossible with PVMs in this scenario.}

It may seem that for purposes of contextuality even the proposition $(\exists \text{ 2-joint}, \exists \text{ 3-joint})$ is of no interest, but there is a subtlety involved here: one is only
considering whether 2-joints or a 3-joint exist for the set $\{M_1, M_2, M_3\}$. Since the statistics that is of relevance for Specker's scenario is the pairwise statistics \cite{LSW, KG}, 
one also needs to consider whether a given choice of 2-joints, $\{M_{12}, M_{23}, M_{31}\}$, admits a 3-joint, i.e., the proposition $(\exists \text{ 3-joint}|\text{ a choice of 2-joints})$ or its negation $(\nexists \text{ 3-joint}|\text{ a choice of 2-joints})$. 
The four possible conjunctions are: 

\begin{itemize}
 \item $(\exists \text{ 2-joint}, \exists \text{ 3-joint})\bigwedge(\exists \text{ 3-joint}|\text{ a choice of 2-joints}),$
 \item $(\exists \text{ 2-joint}, \exists \text{ 3-joint})\bigwedge(\nexists \text{ 3-joint}|\text{ a choice of 2-joints}),$
 \item $(\exists \text{ 2-joint}, \nexists \text{ 3-joint})\bigwedge(\exists \text{ 3-joint}|\text{ a choice of 2-joints}),$
 \item $(\exists \text{ 2-joint}, \nexists \text{ 3-joint})\bigwedge(\nexists \text{ 3-joint}|\text{ a choice of 2-joints}).$
\end{itemize}

Of these, the first conjunction rules out the possibility of KS-contextuality, so it is not of interest for the present purpose. The third conjunction is false since the existence of a 3-joint for 
a given choice of 2-joints would also imply the existence of a 3-joint for the three measurements, hence contradicting the fact that these admit no 3-joints. Thus the two remaining
conjunctions of interest are: 

\begin{itemize}
 \item \emph{Proposition 1}:\\$(\exists \text{ 2-joint}, \exists \text{ 3-joint})\bigwedge(\nexists \text{ 3-joint}|\text{ a choice of 2-joints})$,
 \item \emph{Proposition 2}:\\$(\exists \text{ 2-joint}, \nexists \text{ 3-joint})\bigwedge(\nexists \text{ 3-joint}|\text{ a choice of 2-joints})$\\
$\Leftrightarrow (\exists \text{ 2-joint}, \nexists \text{ 3-joint})$.
\end{itemize}

These two possibilities lead to the following propositions:

\begin{itemize}
 \item \emph{Weak}: $(\exists \text{ 2-joint})\bigwedge(\nexists \text{ 3-joint}|\text{ a choice of 2-joints})$,
 \item \emph{Strong}:\\$(\exists \text{ 2-joint})\bigwedge(\nexists \text{ 3-joint}|\text{ for all choices of 2-joints})$\\
$\Leftrightarrow (\exists \text{ 2-joint}, \nexists \text{ 3-joint})$.
\end{itemize}
 
where \emph{Weak} $\Leftrightarrow$ \emph{Proposition 1} $\bigvee$ \emph{Proposition 2}, and \emph{Strong} $\Leftrightarrow$ \emph{Proposition 2}. The proposition \emph{Weak} relaxes 
the requirement of proposition \emph{Strong} that the three measurements should themselves be incompatible to only the requirement that there exists a choice of 2-joints that do not admit a 3-joint.
Obviously, under \emph{Strong}, there exists no 3-joint for all possible choices of 2-joints: \emph{Strong} $\Rightarrow$ \emph{Weak}.\footnote{
Note that for the case of PVMs, only the conjunction $(\exists \text{ 2-joint}, \exists \text{ 3-joint})\bigwedge(\exists \text{ 3-joint}|\text{ a choice of 2-joints})$
makes sense and that it is, in fact, equivalent to the proposition $(\exists \text{ 2-joint}, \exists \text{ 3-joint})$ since there is no ``choice of 2-joints'' available: the 2-joints,
if they exist, are unique and admit a unique 3-joint (cf. Theorem \ref{uniqueness}). Consequently, the propositions \emph{Weak} and \emph{Strong} are not admissible for PVMs.}

\subsection{Comment on Ref. \cite{FyuOh} vis-\`a-vis Ref. \cite{KG}}

In Ref. \cite{KG}, contextuality---in the generalized sense of Spekkens \cite{Spe05} and by implication in the Kochen-Specker sense---was shown keeping in mind the proposition \emph{Strong}, i.e., requiring that
the three measurements $\{M_1,M_2,M_3\}$ are pairwise jointly measurable but not triplewise jointly measurable. This was in keeping with the approach adopted in Ref. \cite{LSW}, where the construction 
used did not violate the LSW inequality \cite{LSW, KG}. Indeed, as shown in Theorem 1 of Ref. \cite{KG}, the construction used in Ref. \cite{LSW} could not have produced a violation because
it sought a state-independent violation. 

In Ref. \cite{FyuOh}, the authors---under \emph{Proposition 1}---use the construction first obtained in \cite{KG} to show a higher violation of the LSW inequality than reported 
in \cite{KG}. It is easy to check that the construction in Ref. \cite{KG} recovers the violation reported in Ref. \cite{FyuOh} when
the proposition \emph{Strong} is relaxed to the proposition \emph{Weak}: the expression for the quantum probability of anticorrelation in Ref.\cite{KG} is given by

\begin{equation}\label{anti}
R_3^Q=\frac{C}{6}+(1-\frac{\eta}{3})
\end{equation}
where $C>0$ for a state-dependent violation of the LSW inequality \cite{LSW,KG}. Given a coplanar choice of measurement directions $\{\hat{n}_1,\hat{n}_2,\hat{n}_3\}$, and $\eta$ satisfying $\eta_l<\eta\leq\eta_u$, the optimal value of $C$
---denoted as $C^{\{\hat{n}_i\},\eta}_{\max}$---is given by 
\begin{eqnarray}\label{Cmax}\nonumber
 &&C^{\{\hat{n}_i\},\eta}_{\max}=2\eta\\&+&\sum_{(ij)}\left(\sqrt{1+\eta^4(\hat{n}_i.\hat{n}_j)^2-2\eta^2}-(1+\eta^2 \hat{n}_i.\hat{n}_j)\right).
\end{eqnarray}

For trine measurements, $\hat{n}_i.\hat{n}_j=-\frac{1}{2}$ for each pair of measurement directions, $\{\hat{n}_i,\hat{n}_j\}$. Also, $\eta_l=\frac{2}{3}$
and $\eta_u=\sqrt{3}-1$. $\eta>\eta_l$ ensures that the three measurements corresponding to $\{\hat{n}_1,\hat{n}_2,\hat{n}_3\}$ do not admit a 3-joint while
$\eta\leq\eta_u$ is necessary and sufficient for 2-joints to exist: that is, $\eta_l<\eta\leq\eta_u$ corresponds to the proposition \emph{Strong}, $(\exists \text{ 2-joint}, \nexists \text{ 3-joint})$.
On relaxing the requirement $\eta_l<\eta$, we have $0\leq\eta\leq\eta_u$. This allows room for the proposition $(\exists \text{ 2-joint}, \exists \text{ 3-joint})$ when
$0\leq\eta\leq \eta_l$.

The quantity to be maximized is the quantum violation: $R_3^Q-(1-\frac{\eta}{3})=\frac{C}{6}$. Substituting the value $\hat{n}_i.\hat{n}_j=-\frac{1}{2}$ in Eq. (\ref{Cmax}), the quantum probability of anticorrelation from Eq. (\ref{anti}) for trine measurements is given by:
\begin{equation}
 R_3^Q=\frac{1}{2}+\frac{\eta^2}{4}+\frac{1}{2}\sqrt{1-2\eta^2+\frac{\eta^4}{4}},
\end{equation}
which is the same as the bound in Eq. (11) in Theorem 3 of Ref. \cite{FyuOh}. The quantum violation is given by:
\begin{equation}
 R_3^Q-(1-\frac{\eta}{3})=-\frac{1}{2}+\frac{\eta}{3}+\frac{\eta^2}{4}+\frac{1}{2}\sqrt{1-2\eta^2+\frac{\eta^4}{4}}.
\end{equation}

In Ref. \cite{KG}, this expression was maximized under the proposition \emph{Strong} ($\eta_l<\eta\leq \eta_u$)
and the quantum violation was seen to approach a maximum of $0.0336$ for $R_3^Q\rightarrow0.8114$ as $\eta\rightarrow \eta_l=\frac{2}{3}$. In Ref. \cite{FyuOh}, the same expression 
was maximized while relaxing proposition \emph{Strong} to proposition \emph{Weak} (allowing $\eta\leq\eta_l$) and the maximum quantum violation was seen to be $0.0896$
for $R_3^Q=0.9374$ and $\eta\approx 0.4566$.

Another comment in Ref. \cite{FyuOh} is the following: 

\emph{``Interestingly, there are three observables that are not triplewise jointly measurable but cannot violate LSW's inequality no matter
how each two observables are jointly measured.''} 

That is, \emph{Strong} $\nRightarrow$ Violation of LSW inequality. Equally, it is also the case that \emph{Weak} $\nRightarrow$ Violation of LSW inequality.
Neither of these is surprising given the discussion in this note. In particular, note the following implications ($0\leq\eta\leq 1$): 

\begin{enumerate}
 \item Violation of LSW inequality, i.e., $R_3^Q>1-\frac{\eta}{3}$ $\Rightarrow$ Violation of KS inequality, i.e., $R_3^Q>\frac{2}{3}$,
 \item Violation of KS inequality, i.e., $R_3^Q>\frac{2}{3}$ $\Rightarrow$ \emph{Weak}: $(\exists \text{ 2-joint})\bigwedge(\nexists \text{ 3-joint}|\text{ a choice of 2-joints})$,
 \item \emph{Strong} $\Rightarrow$ \emph{Weak}.
\end{enumerate}

Therefore, \emph{Weak} is a necessary condition for a violation of the LSW inequality. It can be satisfied either under \emph{Proposition 1} (as done in \cite{FyuOh}) or under 
\emph{Proposition 2} (or \emph{Strong}, as done in \cite{KG}).

\section*{Joint measurability structures}

I end this note with a comment on the result proven in Ref. \cite{KHF}, where it was shown constructively that any conceivable joint measurability structure for a set of $N$ observables is realizable via
binary POVMs. With regard to contextuality, this result proves the admissibility in quantum theory of contextuality scenarios that are not realizable with PVMs alone. This should be easy to see, specifically, from the 
example of Specker's scenario, where PVMs do not suffice to demonstrate contextuality, primarily because they possess a very rigid joint measurability structure dictated by pairwise commutativity
and their joint measurements are unique (Theorem \ref{uniqueness}). If one can demonstrate contextuality given the scenarios obtained from more general joint measurability structures then a relaxation of
a sort similar to the case of Specker's scenario (from \emph{Strong} to \emph{Weak}) will also lead to contextuality. In this sense, an implication of the result of Ref. \cite{KHF} is that it allows 
one to consider the question of contextuality for joint measurability structures which admit no PVM realization in quantum theory on account of Theorem \ref{uniqueness}.

In particular, for PVMs, \emph{pairwise compatibility} $\Leftrightarrow$ \emph{global compatibility} because commutativity is a necessary and sufficient criterion for compatibility. On the other hand, POVMs allow for 
a failure of the implication \emph{pairwise compatibility} $\Rightarrow$ \emph{global compatibility} because pairwise compatibility is not equivalent to pairwise commutativity for POVMs:
\emph{pairwise commutativity} $\Rightarrow$ \emph{pairwise compatibility}, but not conversely.

\section{Conclusion} I hope this note clarifies issues that may have escaped analysis in Refs. \cite{LSW,KG,KHF,FyuOh}. In particular, the logical relationship between admissible joint measurability structures 
and the possibility of contextuality should be clear from the discussion here.
 
\section{Acknowledgment}
I would like to thank Sibasish Ghosh and Prabha Mandayam for comments on earlier drafts of this article.


\begin{thebibliography}{}
\bibitem{LSW} Y.\ C.\ Liang, R.\ W.\ Spekkens, H.\ M.\ Wiseman, ``Specker's parable of the overprotective seer: A road to contextuality, nonlocality and complementarity''. Phys. Rep. {\bf 506}, 1 (2011).
\bibitem{KG} R.\ Kunjwal,  S.\ Ghosh, ``Minimal state-dependent proof of measurement contextuality for a qubit," arXiv:1305.7009 (2013).
\bibitem{KHF} R.\ Kunjwal, C.\ Heunen, T.\ Fritz, ``All joint measurability structures are quantum realizable,'' arXiv:1311.5948 (2013).
\bibitem{FyuOh} Sixia Yu, C.H. Oh, ``Quantum contextuality and joint measurement of three observables of a qubit,'' arXiv:arXiv:1312.6470 (2013).
\bibitem{Spe05} R.\ W.\ Spekkens, ``Contextuality for preparations, transformations, and unsharp measurements'',
Phys. Rev. A {\bf 71}, 052108 (2005).
\bibitem{heinosaari} T. Heinosaari, D. Reitzner, and P. Stano, Found. Phys. {\bf 38}, 1133 (2008). Or, arXiv:0811.0783 [quant-ph].
\end{thebibliography}
\end{document}